\documentclass[apj]{emulateapj}
\usepackage{graphicx}
\usepackage{amssymb}
\usepackage{amsmath}
\usepackage{natbib}
\usepackage{rotating}
\usepackage{subfigure}
\usepackage{mathtools}

\def\beq{\begin{equation}}
\def\eeq{\end{equation}}
\def\beqa{\begin{eqnarray}}
\def\eeqa{\end{eqnarray}}
\def\bseq{\begin{subequations}\begin{eqnarray}}
\def\eseq{\end{eqnarray}\end{subequations}}

\def\uvp{{{$uv$\ plane\ }}}
\def\kperp{k_{\bot}}
\def\kpar{k_{\|}}
\def\k{{\bf k}}
\def\ka{{\k_{\alpha}}}

\def\ks{{$k$-space\ }}

\def\p{{\bf p}}
\def\An{{\textsf{\textbf{A}}}}
\def\Ea{{\textsf{\textbf{E}}_{\alpha}}}
\def\Eh{{\textsf{\textbf{D}}}}
\def\Eha{\Eh_{\alpha}}

\def\M{{\textsf{\textbf{M}}}}

\def\B{{\textsf{\textbf{B}}}}
\def\Bv{{\tilde{\B}}}
\def\An{{\textsf{\textbf{A}}}}

\def\m{{\bf{m}}}
\def\mp{{\m'}}
\def\u{{\bf{u}}}

\def\uf{{\bf{u}}}
\def\v{{\bf{v}}}

\def\p{{\bf{p}}}
\def\pa{{p_{\alpha}}}
\def\pe{{\p_{\rm e}}}
\def\m{{\bf{m}}}
\def\I{{I}}

\def\Sv{{\textsf{\textbf{S}}}}

\def\sky{{\theta}}

\def\Ih{{\hat{\I}}}

\def\F{{\textsf{\textbf{F}}}}

\def\r{{\bf{r}}}

\def\b{{\bf{b}}}
\def\m{{\bf{m}}}

\def\r{{\bf{r}}}

\begin{document}
\title{Four Fundamental Foreground Power Spectrum Shapes for 21~cm Cosmology Observations}
%\title{The Epoch of Reionization Mode-Mixing Foreground and a Quantitative Framework for Comparing HERA Instruments}
%\title{Analysis of the Epoch of Reionization Mode-Mixing Foreground and a Framework for Quantitative Comparison of HERA Instruments}
%\title{EoR Foreground Mode-Mixing:  a Framework for Removing Mode-Mixing Contamination and Quantitatively Comparing HERA Instruments}

\author{Miguel F. Morales\altaffilmark{1}, Bryna Hazelton\altaffilmark{1}, Ian Sullivan\altaffilmark{1}, Adam Beardsley\altaffilmark{1}}
%\email{mmorales@phys.washington.edu}
\altaffiltext{1}{University of Washington, Seattle, 98195}

\begin{abstract}

Contamination from instrumental effects interacting with bright astrophysical sources is the primary impediment to measuring Epoch of Reionization and BAO 21~cm power spectra---an effect called mode-mixing. In this paper we identify four fundamental power spectrum shapes produced by mode-mixing that will affect all upcoming observations. We are able, for the first time, to explain the wedge-like structure seen in advanced simulations and to forecast the shape of an `EoR window'  that is mostly free of contamination. Understanding the origins of these contaminations also enables us to identify calibration and foreground subtraction errors \textit{below} the imaging limit, providing a powerful new tool for precision observations.

%Astrophysical foregrounds easy to subtract
%Mode mixing from throws purely angular power into line of sight
%Extending the work of Liu and Datta we present a comprehensive description of the mode mixing foreground.
%Describe the origin of the shapes seen in simulations, and provide a mathematical framework for calculating them.
%Provides insight into how to identify the different kinds of errors in the observed power spectra.
%Suggests a way of quantitatively comparing different HERA analysis and instrumental approaches.
\end{abstract}

\maketitle

\section{Introduction}

Observations of redshifted 21~cm emission from neutral hydrogen have the potential to reveal the process of reionization and provide an important new tool for observational cosmology (see \citealt{Furlanetto:2006p341} and  \citealt{Morales:2010p4786} for recent reviews). A number of experiments are currently under construction to observe the 21 cm power spectrum, including LOFAR (LOw Frequency ARray\footnote{http://www.lofar.org/}), PAPER (Precision Array for Probing the Epoch of Reionization\footnote{http://astro.berkeley.edu/~dbacker/eor/}), the MWA (Murchison Widefield Array\footnote{http://www.mwatelescope.org/}), and CHIME (Canadian Hydrogen Intensity Mapping Experiment\footnote{http://www.physics.ubc.ca/chime/}). The primary challenge for these observations is removing contamination from the bright astrophysical foregrounds interacting with instrumental effects and foreground mis-subtractions.

The astrophysical foregrounds are up to five orders of magnitude stronger than the EoR signal, and initial  studies of the 21~cm foreground concentrated on identifying all of the potentially offending sources (e.g., \citealt{DiMatteo:2002p2793,Oh:2003p2809,DiMatteo:2004p2755,Gnedin:2004p2873,Santos:2005p2174,McQuinn:2007p220}; reviewed by \citealt{Morales:2010p4786}). The general consensus is that all known astrophysical foregrounds are either spectrally smooth or  at known  editable frequencies (e.g., galactic radio recombination lines), and none of them mimics the spherical symmetry of the EoR's redshifted emission line. The spectrally smooth astrophysical foreground emission dominates at small $\kpar$ (line-of-sight wave numbers), but quickly falls below the EoR signal strength at higher $\kpar$ values. Conceptually, purely angular modes are dominated by the foregrounds, but the EoR signal can be observed in the line-of-sight (frequency) modes \citep{Zaldarriaga:2004p227,Morales:2004p803,Jelic:2008p4524,Wang:2006p503,Harker:2009p4243}.

The difficulty is that no instrument is perfect, and small instrumental and observational effects can throw the strong angular foregrounds into the frequency ($\kpar$) direction. Effects include chromatic side lobes from sources producing line-of-sight ripples \citep{Bowman:2009p4044,Liu:2009p4716}, small calibration errors leading to mis-subtraction of the chromatic PSFs \citep{Datta:2010p4788}, Faraday rotation of the galactic synchrotron beating with polarization mis-calibrations \citep{Geil:2011p4901}, chromatic primary beams, and numerous other effects. Collectively these are called mode-mixing foregrounds, and have been the primary focus of the EoR foreground subtraction community for the past few years. 

Precision simulations by \citet{Datta:2010p4788} have shown a very distinctive wedge shape in the $\kpar$ vs.\ $\kperp$ power spectrum for three types of mode-mixing contamination---small amplitude errors in bright source removal, small position errors in the bright source removal, and small calibration errors. However, the origin of the wedge shape is not clear. In this paper we expand on the work by \cite{Liu:2011p4789} and \cite{Vedantham:2011p4902} to develop a mathematical framework for subtractive foreground removal and mode-mixing contamination in \S \ref{FrameworkSec}. In \S \ref{ModelSec} \& \S \ref{CalSec} we then explore four mode-mixing signatures and develop a quantitative and qualitative understanding of their origin and why they are fundamental to the measurement process. We conclude in \S \ref{DisSec} by discussing how our new understanding will allow us to calibrate and identify foreground errors \textit{below} the imaging limit.

As the first installment in an informal series of papers on mode-mixing, this paper  concentrates on fundamental contamination common to all observations with subsequent papers exploring array-layout dependent effects and new statistical methods for mitigating these instrumental contaminations.

\section{Framework}
\label{FrameworkSec}

Starting with the notation of \cite{Liu:2011p4789} the band powers $\pa$ from the measured data ($\m$) can be determined with a simplified version of their Equation 2
\beq
\pa = \m^T\Ea\m,
\eeq
where $\Ea$ is the power spectrum analysis and foreground subtraction, we have used the fact that interferometric data is naturally zero mean, and we have omitted the foreground bias term (in essence it is the bias term we are interested in calculating). For this paper it will be easier to split the matrix operator into two symmetric halves 
\beq
\Ea = \Eha^T\Eha,
\eeq
each half describing the linear analysis and foreground subtraction pipeline. The processed data for each band $\mp(\ka)$ (the $\k$-space pixels for which $|\k|$ falls in the spherical shell of the band $\alpha$) is then given by
\beq
\mp(\ka) = \Eha(\ka,\v;\alpha)\m(\v)
\label{Esimpledata}
\eeq
and the band power by
\beq
\pa = \mp_\alpha^T\mp_\alpha = (\m^T\Eha^T)(\Eha\m).
\eeq
Throughout we will use the operator notation ${\textsf{\textbf{A}}}({\bf a},\b; x)$ to signify a transformation from coordinate vector $\b$ to coordinate vector ${\bf a}$ that depends on parameters $x$, so the operator in Equation \ref{Esimpledata} is interpreted as transforming the data from the raw visibilities $\v$ to the $k$-space pixels within a band $\ka$, which depends on the band $\alpha$ we are interested in. A simplified version of a typical analysis pipeline would be
\beq
\Eha(\ka,\v) = \Sv(\ka,\k)\ \F_{\rm 1D}(\k,\uf)\ \Bv^T\!(\uf,\v)
\eeq
where the visibility data is gridded ($\Bv^T$) to the \uvp as a function of frequency ($uv$ vs.\ frequency coordinates denoted by $\u$), the frequency direction is Fourier transformed along the line of sight (frequency) direction and the coordinates are mapped to cosmological wavenumbers $\k$, and the wavenumbers associated with the band power in question are selected ($\Sv$). 

%\cite{Liu:2011p4789} considered multiplicative foreground removal and how to mitigate biases  introduced by the foreground removal steps. In this paper we are going to build off their work and consider the more commonly discussed subtractive foreground removal methods.

While most astrophysical foregrounds are smooth in frequency and thus separable from the nearly spherical EoR signal in principle \citep[all emission near $\kpar = 0$, ][]{Morales:2006p147,McQuinn:2006p222}, the effect of chromatic instrument responses, imperfect foreground models, and/or imperfect calibration conspire to throw this nearly purely angular foreground power into the line-of-sight $\kpar$ direction, masking the faint EoR signal. Here we label the power spectra due to various mode mixing terms as $\pe(\k)$, which is defined as the square of the data residuals $\r(\k)$ in $k$-space ($\pe = |\r|^2$). We can formally calculate the shapes of these mode-mixing power spectra by considering the difference between the observed and subtracted foregrounds. Looking at the residuals for a simplified software holography chain we have 
\beqa
 \r(\k) =  \F_{\rm 1D}(\k,\uf)\Bv^T(\uf,\v)\M( \v, \uf)\F(\uf,\sky) \I(\sky) \hspace{10pt} \nonumber \\
 -\  \F_{\rm 1D}(\k,\uf) \Bv^T(\uf,\v) \B( \v, \uf) \F(\uf,\sky) \Ih(\sky), \hspace{10pt}
\label{Eresidual}
\eeqa
or annotated  
\beqa
 \r(\k) = \overbrace{
 \underbrace{\F_{\rm 1D}(\k,\uf) \Bv^T(\uf,\v)}_\textrm{analysis}\ 
 \underbrace{\M(\v,\uf)}_\textrm{instrument} \F(\uf,\sky) 
 \underbrace{I(\sky)}_{\mathclap{\substack{\textrm{true}\\ \textrm{foreground}}}}
 }^\textrm{observed foreground}\nonumber \\
\nonumber \\ 
 -\ \overbrace{
 \underbrace{\F_{\rm 1D}(\k,\uf)\ \Bv^T(\uf,\v)}_\textrm{analysis}\ 
 \underbrace{\B(\v,\uf)}_{\mathclap{\substack{\textrm{model}\\ \textrm{instrument}}}}\ \F(\uf,\sky) 
 \underbrace{\Ih(\sky)}_{\mathclap{\substack{\textrm{model}\\ \textrm{foreground}}}}.
 }^\textrm{subtracted foreground} \nonumber
\eeqa
The first line of Equation \ref{Eresidual} shows the observed foreground as the true foreground $I$  measured by the instrument $\M$ and run through a holographic analysis $\F\Bv^T$. The second line is the subtracted foreground, with the model foreground $\Ih$ and model instrument $\B$ replacing the true foreground and instrument. The analysis portion of how visibilities are transformed into $k$-space measurements are usually the same for both the measurement and the model (they can differ for computational reasons), so we can introduce an analysis operator $\An(\k,\v) = \F\Bv^T$. 

In general the residual foreground contamination is due to an admixture of model errors ($\Ih \ne I$) and calibration errors ($\M \ne \B$), but it is instructive to look at these sources of errors independently
\beqa
\r_{\rm M} = \An\B(I - \Ih) \hspace{.25 cm} \textrm{model error, calibration correct\ \ }\label{ModelError}\\
\r_{\rm C} = \An(\M-\B)I \hspace{.25 cm}  \textrm{calibration error, model correct.}\label{CalError}
\eeqa
Formally our work is now done. We can parametrize the kinds of errors we make, e.g.\ amplitude and or position model errors in Equation \ref{ModelError}, and calculate the associated power spectra $\p_{\rm e}(\k)$ for our instrument.

 However, it is useful to qualitatively identify the origin of these shapes. In \S \ref{ModelSec} \& \ref{CalSec} we identify four fundamental residual power spectrum shapes that will be seen by all upcoming instruments and describe their origins. In this paper we concentrate specifically on the contamination that will be seen by any upcoming observation, independent of the array layout. In subsequent papers we explore the effects of array layout and non-uniform sampling of the $uv$ plane (Hazelton et al.\ in preparation), and more advanced statistical techniques for removing the instrumental effects. For this paper we limit ourselves to effects related to fundamental information loss on the antenna scale.

\section{Model Error Shape and the Foreground Wedge}
\label{ModelSec}

First we consider the effect of foreground model errors. In this case we assume the calibration of the instrument is perfect---both the complex gain and beam shape of each antenna is known exactly---but the subtracted foreground model is not equal to the true sky (Equation \ref{ModelError}). Typical model errors include mistakes in the amplitude and location of compact sources, errors in the amplitude and shape of diffuse emission, and polarization errors for both compact and diffuse sources. Both thermal noise and systematics contribute to foreground model errors, and the PSF sidelobes often make it impossible to disentangle the true brightness and locations of the sources in the field. Accurately determining the foregrounds is particularly difficult in the confusion limit, and \cite{Liu:2009p4716} and \cite{Bowman:2009p4044} have done excellent work on improving the accuracy of the foreground models in the confusion limit.

In Figure 10 of \cite{Datta:2010p4788} random errors in the foreground model generated a very distinctive 2D \ks ``wedge''  which has now been seen in observations with the MWA prototype (Bernardi, et al.\ in preparation). As we will explain in this section, this wedge is due to the chromatic instrument response and inherent information loss.

Any interferometric measurement suffers information loss on the scale of one antenna. An antenna sums the electric field across its surface into a single voltage signal which is cross-correlated with the integrated electric field from a second antenna to form a visibility. In this process the variations in the electric field correlation across the antenna surface (e.g.\ from a source far from field center) are lost. Equivalently, in the \uvp one visibility is formed by integrating the true $uv$ correlation with the beam pattern of that antenna pair $\B( \v, \uf)$ over a small frequency channel as shown in Equation \ref{Eresidual}.  Modern array simulators use this integral of the \uvp to predict the visibilities including direction dependent gain patterns. \cite[For a conceptual introduction to interferometric measurement, please see Chapter 3 of][]{Morales:2010p4786}. The first step of any analysis is to reconstruct an estimate of the \uvp $\Ih(\uf)$ by gridding the visibilities ($\Bv^T$ in Equation \ref{Eresidual}). However, $\B$ is not invertible and the the operation $\Bv^T\B$ does not recreate the true \uvp distribution of the sky.  

Various statistical priors can be used to mitigate this information loss, and deconvolution is a powerful technique for reconstructing the contributions to each visibility below the antenna scale. Effectively the statistical prior that most of the flux is due to a list of sources (or CLEAN components) allows their contribution to each visibility to be determined accurately despite the smearing due to the antenna integration in the $uv$ plane.  
%[Aside: there is often concern over the non-linearity of deconvolution and the resulting non-Gaussian distribution of errors. Not only are the estimated fluxes and positions slightly wrong, the distribution of the error in the flux and position are not Gaussian. While this is a real effect, for the model errors we do not require that they be Gaussian, and any errors, Gaussian or not, will produce the same residual wedge in \ki-space. The non-Gaussianity only affects the amplitude and statistics of the wedge.]

However, the EoR signal is below the imaging noise floor for the first generation observatories (S/N less than one per per visibility, or equivalently, phase noise greater than one radian). Thus deconvolving the EoR signal is not possible, and we must inherently work with a `dirty' residual map (this is also true for all CMB measurements and other diffuse power spectrum measurements). Foreground model errors appear as small spurious signals in the EoR dirty map. First we will explore the effect of amplitude errors which exhibit the basic features common to several of our model and calibration error signatures, then we will describe source locations errors in the subsequent sub-section.

\subsection{Amplitude errors}
\label{AmpSec}

\begin{figure}
\begin{center}
\includegraphics[width = \columnwidth]{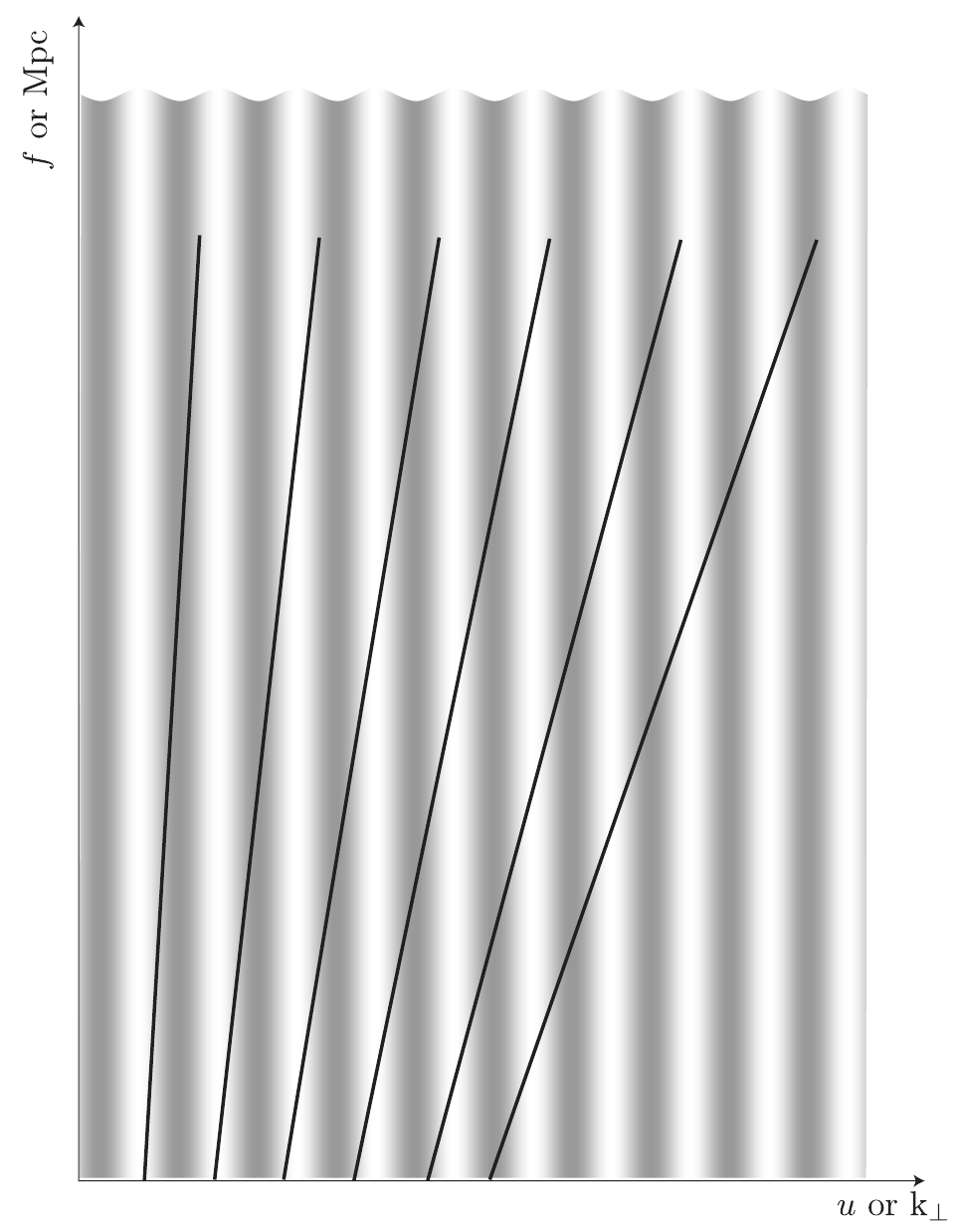}
\caption{This cartoon represents a slice through the three dimensional $\u,f$ space, where $u$ is proportional to $\kperp$ and $f$ is proportional to line-of-sight distance in Mpc. The vertical corrugation shows the spatial-frequency corrugation of a flat spectrum source away from the center of the field (real part shown). The diagonal stripes show the paths of individual antenna baselines as a function of frequency. While the distance between two antennas in wavelengths (or $\kperp$) increases with frequency for all baselines, the mis-alignment angle is larger for longer baselines. It is this increasing mis-alignment of the baselines with the corrugations of a residual source that create the foreground wedge.}
\label{Figuf}
\end{center}
\end{figure}

An error in source brightness will leave a small residual positive or negative source at the location of the true source, and is conceptually the simplest of the contaminations. Figure \ref{Figuf} shows the true correlation of a flat spectrum source far from the field center in the $\u,f$ space (or equivalently $\kperp$ and line-of-sight Mpc) with a few baselines overlaid. The oscillation in $\kperp$ is given by the distance from the field center (quick oscillations for sources near field edge) and for a flat spectrum source amplitude is constant in frequency. The baselines all migrate to larger distances (in wavelengths) with higher frequency, with the slope of the chromatic migration increasing with baseline length.

\begin{figure*}
\begin{center}
\subfigure{ \includegraphics[width = \columnwidth]{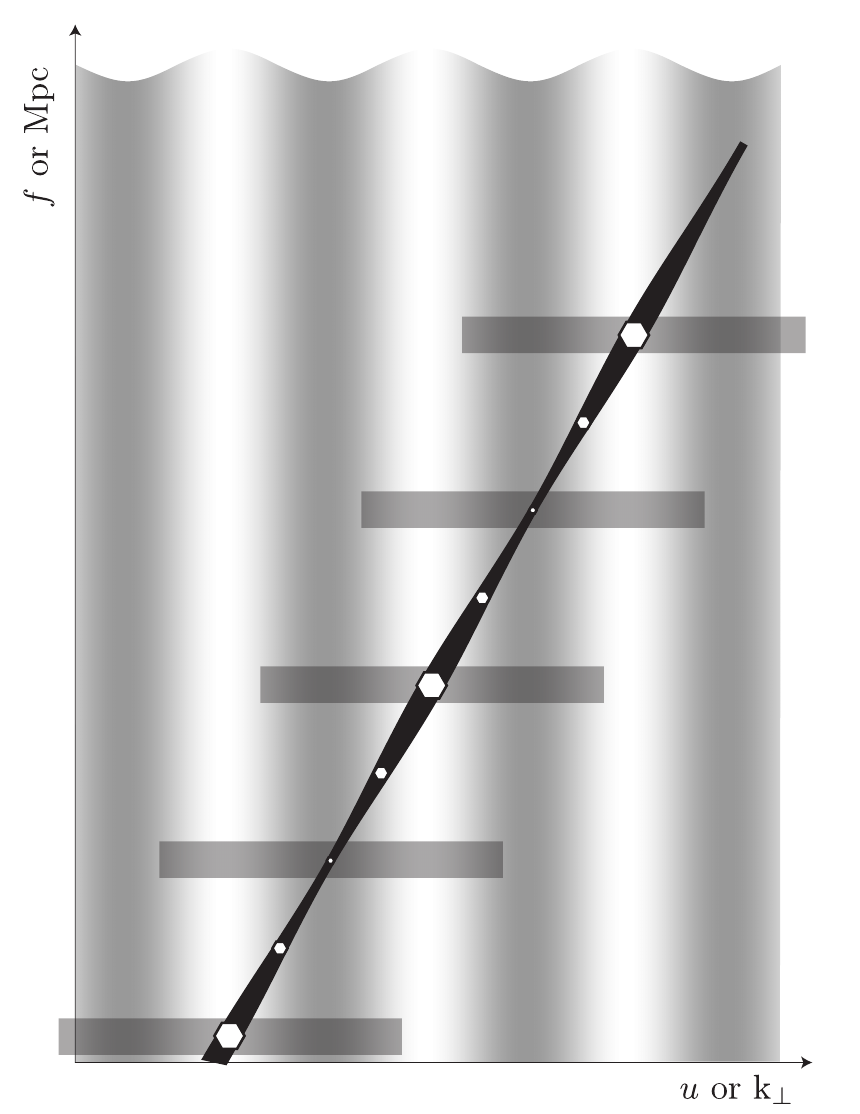}}
\subfigure{ \includegraphics[width = \columnwidth]{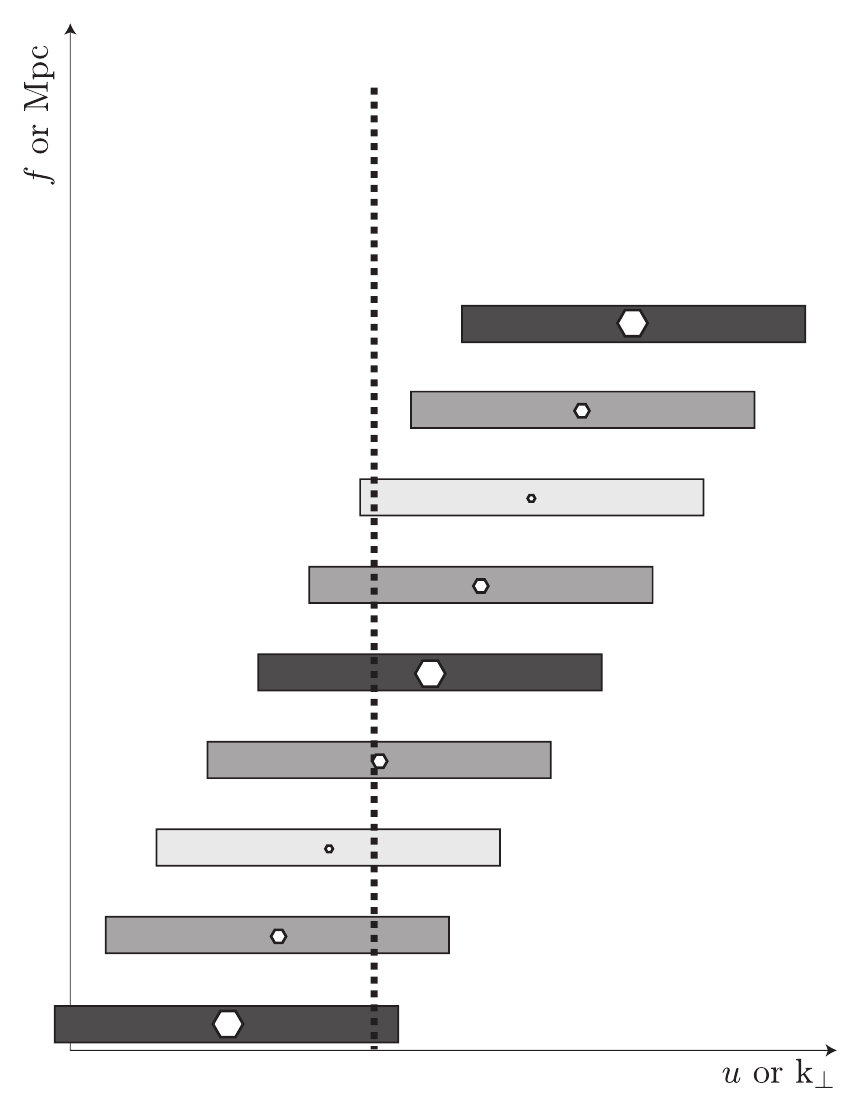}}
\caption{These figures zoom in two a small region of the $\u,f$ plane to study the effect of baseline chromaticity. In the left hand panel, the vertical corrugations from a residual source are shown, with a line showing the measured visibility. The measured visibility is the integral of the true correlation function (grey corrugations) times the antenna beam (grey boxes in left hand figure). The resulting measured visibility wraps as a function of frequency, as indicated by the line width and the size of the hexagons. The right hand figure then shows the result of the gridding process, where the measured visibilities are gridded to the \textit{reconstructed} $\u,f$ plane. Effectively the measured visibility is smeared over a region of $\u$ by the gridding kernel. Light grey indicates a small (real component) reconstructed in that region of $\u,f$, while dark grey indicates a large value. When we Fourier transform in the line-of-sight direction to form the power spectrum cube (Figure \ref{FigWedge}), we will see oscillations in the line-of-sight direction as indicated by the oscillating grey values along the vertical dashed line.}
\label{Figufzoom}
\end{center}
\end{figure*}

Figure \ref{Figufzoom} then zooms in on a small region of Figure \ref{Figuf} to explore the effects of measurement (lefthand panel) and reconstruction (righthand panel) for a single baseline. In the left hand panel the true correlation for the residual source (vertical corrugations) is integrated by the antenna beam (grey rectangles) to form the residual visibility. The measured visibility traces the true correlation as shown by the size of the hexagons and the line width.  (Mathematically this is a sampled convolution in $\u$ of the beam and the true correlation---or the true correlation times the direction dependent gain of that antenna pair.) As expected, the measured visibility oscillates as a function of frequency due to the chromatic increase in baseline length.

The difficulty comes in the reconstruction of an unbiased estimate of the sky. As previously mentioned we have lost information both where there are no baselines (missing spatial frequencies) and on scales smaller than the antenna in the process of measuring the signal.  Because the signal to noise on the EoR signal is much less than one for a single baseline, we must resort to gridding the visibilities to form a dirty map. Gridding with the holographic antenna pattern guarantees an unbiased power spectrum measurement \citep{Morales:2009p4730,Tegmark:1997p2012}, but the exact form of the gridding is not important for this discussion.

The righthand panel of Figure \ref{Figufzoom} shows the effect of gridding for one baseline. The sinusoidal variations in the visibility are spread out by the gridding kernel (boxes), resulting in regions of high and low amplitude represented by the grey scale. This produces oscillations in the line-of-sight direction as seen by the varying amplitude along the dashed line. This oscillation is not in the original source (lefthand panel), but because we lost information on the scale of the antenna unavoidably shows up in our reconstructed signal (right hand panel). When we Fourier tranform in the line-of-sight direction to get into the three dimensional $\k$-space this reconstruction error throws power that was originally purely in the $\kperp$ direction into the $\kpar$ dimension---the hallmark of mode-mixing.

\begin{figure}
\begin{center}
\includegraphics[width = \columnwidth]{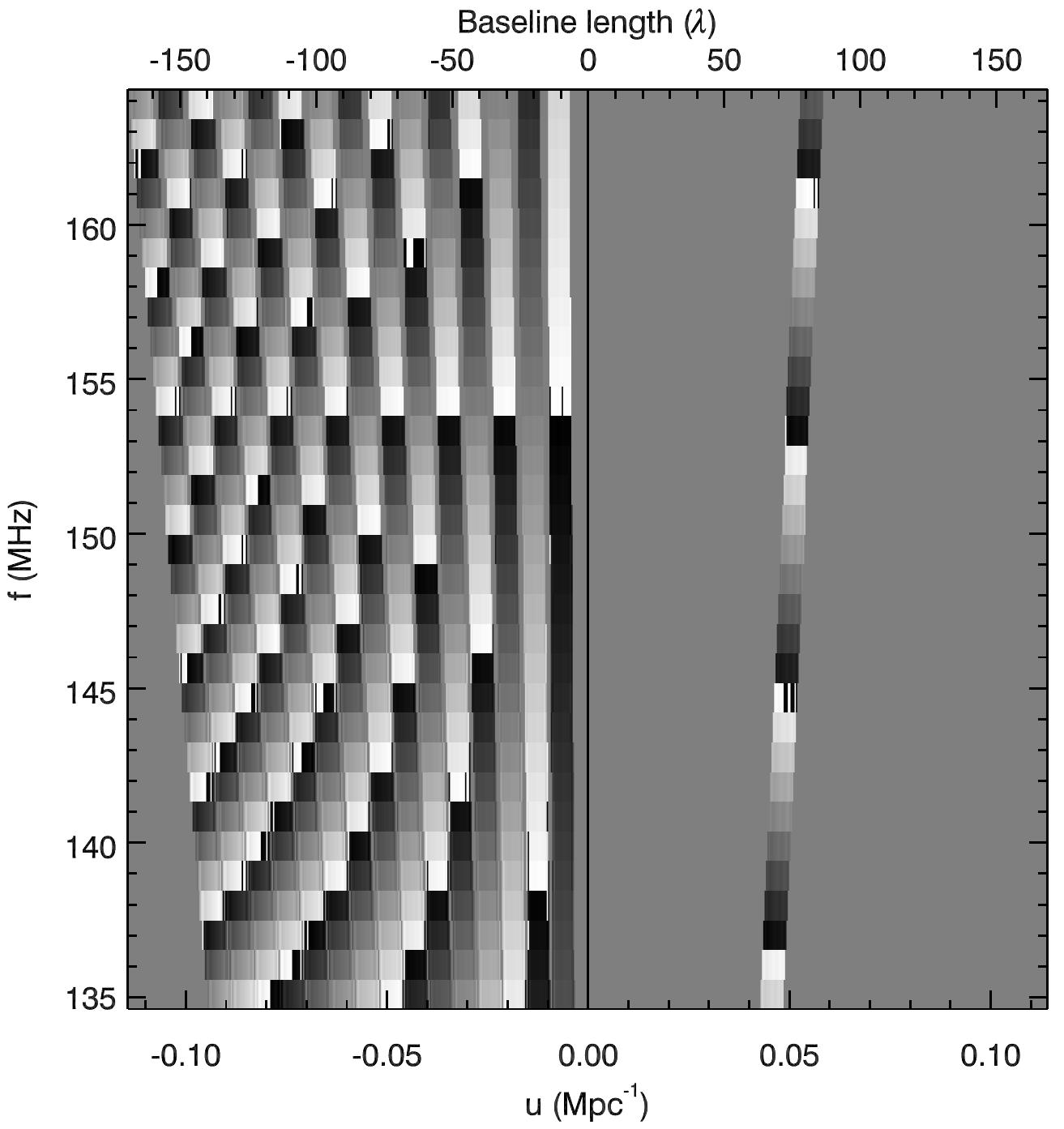}
\caption{This figure represents a $uf$ slice through a precision simulation of a simple array, and illustrates the contamination  described in Figures \ref{Figuf}--\ref{FigWedge}. On the right half of the figure (positive $u$) there is a single baseline in a reconstructed $uf$ plane (analogous to Figure \ref{Figufzoom}b), and on the left side there are numerous baselines of different lengths. The color scale shows the reconstructed phase for a single residual source $14^\circ$ from the field center (black = $-\pi$, white = $+\pi$, grey = 0 phase or no baseline contribution). The phase wrap is clearly seen with increasing frequency in the single baseline on the right, and the increasing rate of phase wrap is seen in the longer baselines on the left.}
\label{lineSim}
\end{center}
\end{figure}

Figure \ref{lineSim} shows a precision simulation similar to the simulations by \cite{Datta:2010p4788} of a single mis-subtracted off-axis point source as seen by a simplified array. In addition to the phase wrap predicted in Figure \ref{Figufzoom}b, the longer baselines produce higher frequency contamination due to their higher chromatic migration angles. Mathematically, the wavelength of the residual source in the $\u$ direction $\lambda_\perp$ is related to the wavelength of the contamination in the line-of-sight $\lambda_{\rm LoS}$ direction by the baseline to wavelength conversion $\u(\lambda) = \u(m)f/c$, giving the relationship
\beq
\frac{\lambda_{\rm LoS}}{\lambda_\perp} = \frac{c}{\u(m)} = \frac{f}{\u(\lambda)}.
\eeq
Fourier transforming in the line-of-sight direction and converting to cosmological coordinates \citep{Morales:2004p803} one obtains, after a little algebra,
\beq
%\k_{||} \approx \sky'(\rm rad)\ \kperp \left(\frac{D_M(z)}{D_H}\frac{E(z)}{2\pi(1+z)}\right).
\k_{||} \approx \sky'(\rm rad)\ \kperp \left(\frac{D_M(z)}{D_H}\frac{E(z)}{(1+z)}\right).
\label{kcont}
\eeq
$\sky'(\rm rad)$ is the vector distance of the residual source from the field center, $\kperp$ is the baseline length in cosmological coordinates, and the cosmological terms in parenthesis range from 
%0.1 
0.63 at a redshift of 1 to 
%0.54 
3.4 at a redshift of 8 \citep[$\Omega_\Lambda$ = 0.728, h = 0.704, flat universe following notation of ][]{Hogg:1999p923}.

\begin{figure}
\begin{center}
\includegraphics[width = \columnwidth]{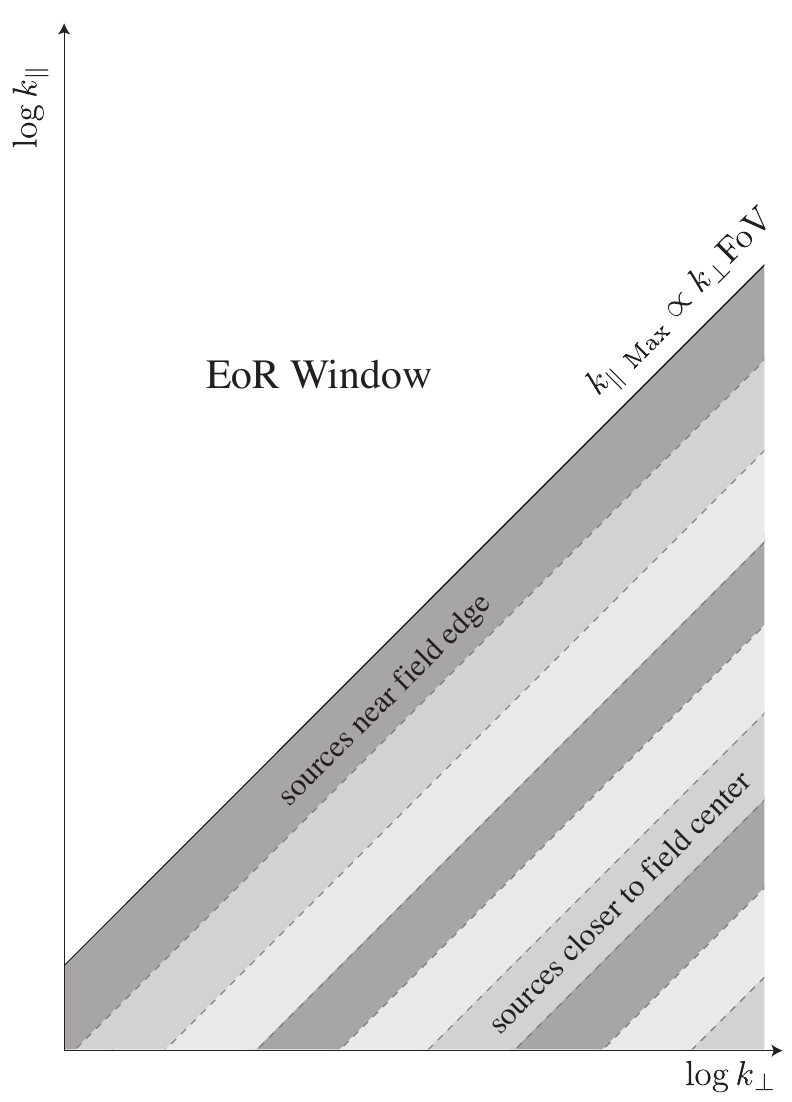}
\caption{This figure shows a slice through the $\kpar$ vs. $\kperp$ space on logarithmic coordinates. The residual contamination described in Figures \ref{Figuf} \& \ref{Figufzoom} appears as a line of contamination along $\kpar \propto \kperp$, with the intercept depending on distance from the phase center. The instrumental field-of-view provides a soft cutoff in the contamination, leaving an EoR window that is free of foreground model errors. The amplitude of the contamination depends on the PSF of the instrument (smoother PSF leads to lower contamination), but in general it will be difficult to remove foreground contamination from within the mode-mixing wedge. For amplitude errors the contamination is uniform in amplitude along lines of $\kpar \propto \kperp$, while the contamination from location errors increases $\sim$linearly with $|k|$.}
\label{FigWedge}
\end{center}
\end{figure}

The contamination in Equation \ref{kcont} is remarkably simple:  a power law of 1 relationship between $\kperp$ and $\kpar$, depending only on the angle of the source from the field center and an order one constant based on the cosmology. Figure \ref{FigWedge} shows how this contamination will appear in the $\kperp$ vs.\ $\kpar$ space. This is an approximate relationship, as only a small length of the line-of-sight oscillation is contributed by one baseline, and the contributions of different baselines (at same $\kperp$) are incoherent. The sum of these short incoherent oscillations at the same wavelength is a strong peak at the associated line-of-sight wavenumber $\kpar$, but will contain power at other wavelengths and the magnitude of the mode-mixing will decrease as more baselines are added (smoother PSF).

There is, however,  a soft limit to the line-of-sight contamination imposed by the instrument field-of-view. The corrugations of the residual sources are band limited by the instrument field-of-view, giving a maximum line-of-sight contamination of
\beq
%\k_{||\ {\rm Max}} \approx \tfrac{1}{2}{\rm FoV(radians)}\ \kperp \left(\frac{D_M(z)}{D_H}\frac{E(z)}{2\pi(1+z)}\right).
\k_{||\ {\rm Max}} \approx \tfrac{1}{2}{\rm FoV(radians)}\ \kperp \left(\frac{D_M(z)}{D_H}\frac{E(z)}{(1+z)}\right).
\eeq
This identifies a convenient EoR window above $\k_{||\ {\rm Max}}$ where foreground model errors will not contaminate the EoR signal (after \citealt{Vedantham:2011p4902}). As we will see in later sections, the $\k_{||\ {\rm Max}}$ line plays an important role in determining the region where the EoR signal can be measured and the relative importance of different contaminations. 

The small residual sources from amplitude errors have a constant magnitude as a function of $\kperp$ (FT of a $\delta$-function), so the amplitude along any dashed line in Figure \ref{FigWedge} is constant. However, different lines in the wedge correspond to sources at different (projected) distances from the phase center so the amplitude can vary perpendicular to the $\k_{||\ {\rm Max}}$ line. The characteristic shape of amplitude errors is a wedge of power below $\k_{||\ {\rm Max}}$ that is constant parallel to the $\kpar \propto \kperp$ diagonal but varies in the orthogonal $\kpar \propto -\kperp$ direction, with errors closer to the field edge appearing higher in the wedge. In the simulations by \cite{Datta:2010p4788}, mis-subtracted sources were scattered uniformly across the image, creating many such diagonal lines for sources at different projected distances from the image center. These many diagonal lines in the power spectrum added together to produce the wedge seen in that work.

\subsection*{Why don't we just...}

The origin of the foreground wedge is the inherent information loss below the antenna scale, coupled with our inability to deconvolve when the signal-to-noise is less than one. However, this is not at all self apparent and it is natural to suggest a number of alternatives. Common ideas include:

\textit{Why don't we just grid in meters instead of wavelengths?} Gridding in meters skews Figures \ref{Figuf} \& \ref{Figufzoom} so the baselines are vertical but the corrugations of the residual source and the line-of-sight Fourier transform are at an angle (which increases with baseline length). The mode-mixing is unchanged by this coordinate transformation---the same effect occurs, just in a skewed coordinate frame. While gridding in meters may be more computationally efficient, the effect described here is independent of the gridding coordinates.

\textit{ Why don't we Direct Fourier Transform the visibilities without gridding?} This seems like the obvious solution as it eliminates the gridding step entirely, but there are two reasons this does not solve our problem. First, the antenna field-of-view changes significantly with frequency (even if all the antennas are identical to each other), and we must be able to insert this into our analysis. One way of doing this is to form maps at different frequencies via DFT, which are then tapered with the primary beam before assembling into an image cube (this is what NVSS did to assemble mosaics). However, multiplying by an image space taper is identical to convolving in the $uv$ plane. This smooths out our visibilities in the $uv$ plane, reproducing the effect seen in the righthand panel of Figure \ref{Figufzoom}. Gridding with the antenna beam is identical to tapering in the image (this duality is the basis of software holography). Any image tapering reproduces this effect.

One can consider performing a full sky DFT and not applying a taper (as one might be tempted to do if the antenna beam was frequency independent), but this turns out to be worse for a power spectrum analysis. Effectively, contributions from a visibility are reconstructed across the full sky, even in areas where the telescope had no sensitivity (e.g. far outside the primary beam). This introduces \textit{bias} into the power spectrum measurement \citep{Tegmark:1997p2012}. Fundamentally we know the contributions to a single visibility came from a region of the $uv$ plane, not a point, and we need to reconstruct this effect. We can do this either by gridding in $uv$ or tapering the image, but not doing either actually biases our result by over emphasizing areas of the sky.

\textit{What if we only include visibilities which exactly overlap in $\u$ when performing the line-of-sight FT?} The first problem is that if the antenna field-of-view changes, the integral of the $uv$ plane that went into making those visibilities differs. This reproduces the  difficulty discussed in the previous question and tapering/gridding is needed to calibrate the data. Even if the antennas are not chromatic, this represents a drastic decrease in the effective collecting area. One cannot design a two dimensional antenna layout that provides the necessary $uv$ coverage without discarding the majority of the baselines (even a regular grid of antennas only has good redundancy along the two grid directions).

\textit{What about deconvolving the EoR signal itself}? This could be done for the SKA or other instruments which will image the EoR and BAO fluctuations. However, deconvolution requires a signal to noise per visibility greater than one (phase noise $\lesssim$ one radian) and the first generations of instruments will not have the sensitivity.

\textit{What if we measured every baseline (at the scale of $\lambda/2$ or better)?} This would eliminate the contamination we have identified, not by removing the wedge but by driving its amplitude to zero. With dense enough baseline coverage the incoherent summing of many line-of-sight oscillations in the righthand panel of Figure \ref{Figufzoom} would push the amplitude to zero. However, this criteria is equivalent to building an instrument with a $\delta$-function PSF. Desirable but impractical.

We believe that the foreground wedge identified here is fundamental. It is a product of the information loss at the antenna scale suffered in making the interferometric measurement, and the need to reconstruct a faithful `dirty' reconstruction of the sky.

\subsection{Location errors}
\label{LocationSec}

Location errors assume that the amplitude of the model source is correct, but the location is slightly wrong---typically by much less than one beamwidth. In the absence of calibration errors, this is equivalent to having two $\delta$-functions with opposite sign very close to one another. After transforming to the $\u,f$ frame of Figures \ref{Figuf} \& \ref{Figufzoom}, the Fourier shift theorem shows the contamination to be the subtraction of two sinusoidal corrugations of slightly different frequency. These subtract perfectly at zero $\kperp$ with the amplitude growing with $\kperp$ as the corrugations slowly creep out of phase. In the limit of offsets much smaller than one beam ($\ll$ one radian of phase at longest baseline), this gives a sinusoidal corrugation whose magnitude  increases approximately linearly with $\kperp$.

We can then repeat the argument of the previous section and Figures \ref{Figuf}--\ref{FigWedge}, the only difference being that the source of the contamination is a linearly increasing function of $\kperp$ instead of a constant. This limits the contamination to the same region as the amplitude errors, with the amplitude of the contamination proportional to $|k|$ along lines of  $\kpar \propto \kperp$. Thus location errors preserve the same EoR window above $\k_{||\ {\rm Max}}$ but with a distinctive linear increase in amplitude along lines of  $\kpar \propto \kperp$. This is the origin of the distinctive wedge shape seen in Figure 10 of \cite{Datta:2010p4788}.

%Location errors---the error should increase with \kperp. The error is a shift that increases with angular frequency, thus the subtraction is worse at large \kperp. Quasi-linear in the small angle approximation, then increases to some maximum value related to the rumble level (approximately the same as no subtraction). However, this limit is for a full pixel off, and the errors are usually much smaller than that.  
%[Amplitude errors are obviously small spurious sources, but offsets are as well. Mathematically they are the true source and the model as opposite $\delta$-functions very near one another. These produce corrugations in \u which have the same amplitude but very slightly different frequencies. Integrated over an antenna beam it gives a small additional contribution to the visibility---the same effect as a small error in the source amplitude for that visibility. This amplitude slowly increases with baseline length. ] 

%Deconvolution can clean below this limit, but cannot deconvolve model errors (otherwise one fixes the model). 
%Band limit is really due to attenuation by the integral B(v,[u,f])
%Loss of information on the scale of the beam.
%SH simply the least biased regridding function.
%We believe the use of a spheriodal in Datta has increased the extent of kpar contamination, due the ringing in the tails of the spheroidal kernel.

%Dig into the analysis, describe transfer function, introduce Figure
%Discuss the origin of the wedge
%Figure showing it's shape.
%Discuss spatial identification of errors in k space
%shape difference between amplitude and location errors.

\section{Calibration Error Shapes}
\label{CalSec}

In this section we consider the effect of calibration errors. We now assume that the foreground model is equal to the true sky, but that the model instrumental response is not equal to the true response ($\B \ne \M$, Equation \ref{CalError}). There are a number of ways the instrumental calibration can be wrong, from simple gain errors (visibility amplitude error) to errors in the antenna FoV or gain profile (wrong integral of the true $uv$ plane) to assuming all antennas are identical (not accounting for antenna-to-antenna variation) to antenna location or survey errors (error in location of a visibility) or full fledged direction and frequency calibration errors. 

To begin with we will concentrate on small frequency-independent errors on a subset of the baselines. Simple gain errors and differences between antennas will be reflected in a few visibilities with the wrong complex amplitude. If most of the visibility calibrations are correct (and the model is correct), the correct visibilities will subtract perfectly leaving only the erroneous component of the miscalibrated baselines. These calibration errors lead to two residual power spectrum shapes---a diffuse component that contaminates nearly all $\k$ modes and a third wedge shape that is similar to the foreground model residuals of \S \ref{ModelSec} but with a distinct functional form.

%Make a figure with many baselines greyed/dotted out, but a few which are black with a band showing where its contribution appears in the uf plane. \label{Figufcal}
\begin{figure}
\begin{center}
\includegraphics[width = \columnwidth]{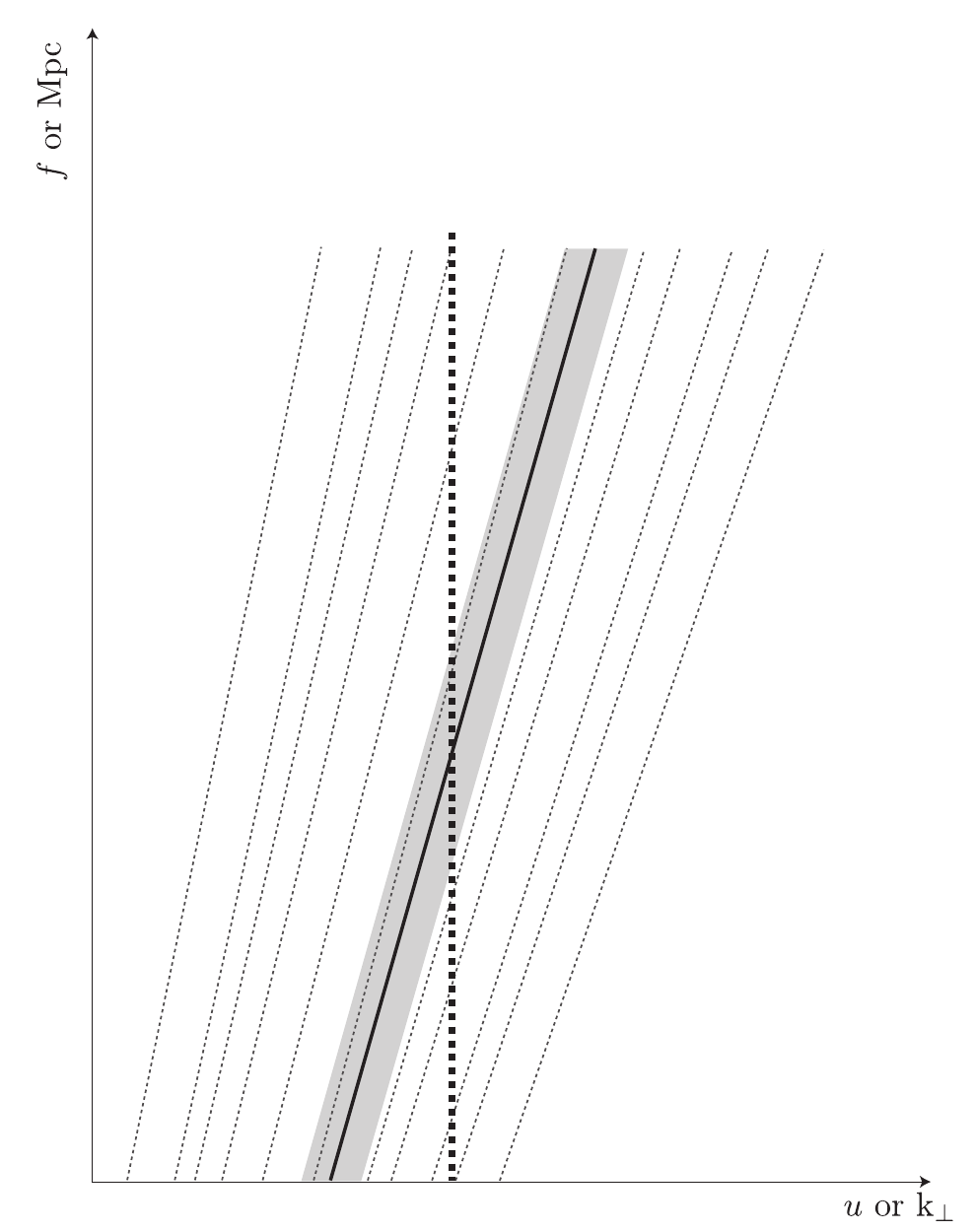}
\caption{This cartoon shows the effect of calibration errors in a portion of the $\u, f$ (or $\kperp$, line-of-sight Mpc) space. The diagonal lines show the paths of a number of visibilities (unequally spaced). The dotted visibilities are correctly calibrated and have subtracted perfectly, but one visibility had a small calibration error so left a residual visibility (solid line). This residual visibility is gridded to the $u,f$ plane by the grey shaded region. After Fourier transforming along the frequency direction (dashed line), this residual visibility produces contamination at all $\kpar$ modes. Effectively it appears as the Fourier transform of a windowed $\delta$-function (see text for details).}
\label{Figuvcal}
\end{center}
\end{figure}

Figure \ref{Figuvcal} shows the effect of this calibration residual, where the dotted visibilities have subtracted perfectly but the black visibility has not. An error in the gain amplitude leaves a residual visibility that is in phase with the underlying foreground signal, while a phase error leaves a residual visibility that is $\pm\pi/2$~rad out of phase. The residual complex visibility is then gridded to the $u,f$ plane as shown by the grey region. When taking a Fourier transform along the line-of-sight direction (indicated by the dashed line) this gain error appears as a windowed $\delta$-function. Effectively the calibration error for a visibility produces a  sharp feature in the frequency direction, producing contamination over a wide range of $\kpar$ modes. 

\begin{figure}
\begin{center}
\includegraphics[width = \columnwidth]{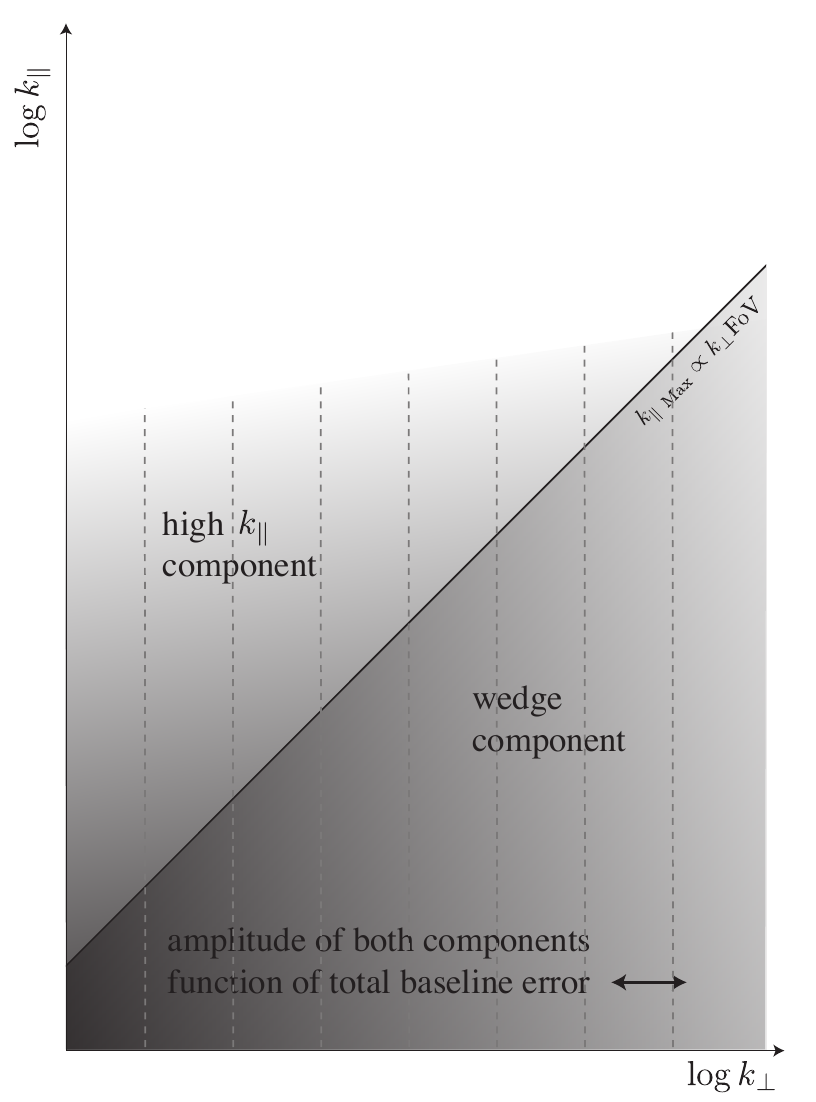}
\caption{This cartoon shows the effect of calibration errors in the $\kpar$ vs.\ $\kperp$ space. The first component reaches to high $\kpar$, contaminating the EoR window. This contamination is windowed (represented by grey scale), and the effectiveness of this windowing depends on the details of the analysis. There is also a wedge component due to calibration errors that is similar to the two foreground wedges, but depends on the total baseline error with a $\sim \kperp$ functional form.  }
\label{Figkcal}
\end{center}
\end{figure}

The contamination in $\kpar$ vs. $\kperp$ space appears as a flat contamination at all $\kpar$ ($\delta$-function FT) times a window function that falls with increasing $\kpar$, as shown in Figure \ref{Figkcal}.  If the visibilities are gridded with the beam pattern $\B(\v,\u)$, the windowing function is a scaled version of the antenna beam pattern---the dashed line in Figure \ref{Figuvcal} traverses the $uv$ beam pattern and is Fourier transformed to produce a scaled version of the angular beam pattern in the $\kpar$ direction. This is the most insidious of the foregrounds detailed in this paper because the contamination extends into the `EoR window.' 

%New figure showing a gradient for kpar and kperp dependent wedge. 
%\label{Figkcal}

%If two  closely spaced visibilities have different errors, they appear as separated $\delta$-functions and will throw power on the scale of the inverse frequency separation between them (the beat frequency). Thus small calibration errors can contaminate a wide swath of $\kpar$ extending throughout the EoR window. 

%The density of the visibility sampling along the line-of-sight has very little effect on this contamination. If there are only 10 visibilities along one line-of-sight, there are only $\sim$10 $\kpar$ modes that can be observed. Random calibration errors throw power over the entire range of line-of-sight wavenumbers available to the instrument. Denser visibility sampling provides a larger range of $\kpar$ modes, but they are all contaminated by errors in calibration. (Due to the steepening of short baselines in $\u,f$, the range of $\kpar$ modes available tends to decrease towards shorter baselines. Sometimes $\kpar$ vs.\ $\kperp$ plots extend beyond the band limit of the instrument, producing what appears to be a second EoR window but is instead an area of the space unexplored by the instrument.)

The calibration errors also produce another wedge shape. The visibilities with calibration errors see sources from across the sky and oscillate with frequency just like in Figure \ref{Figufzoom}. This again produces the same kind of $\k$ space contamination, but because the sky is filled with sources it naturally contains sources from all distances from phase center up to the field-of-view cutoff. The difference is that the contamination is only over the range of $\kperp$ (or $\u$) impacted by that baseline.  For the random calibration errors on all baselines in Figure 11 of \cite{Datta:2010p4788}, this produces a wedge of contamination below $\k_{||\ {\rm Max}}$ with a functional form that follows the baseline distribution. Unlike the the foreground model errors which are functions of the diagonal $\kpar = \kperp$ and $\kpar = -\kperp$ axes, the calibration error is a function of the baseline distribution and is  $\sim$parallel to $\kperp$. 

%We can now understand the contamination generated random calibration errors on all baselines seen in Figure 11 of \cite{Datta:2010p4788}. There is a wedge due to the residual visibilities and a component that extends to high $\kpar$, both of which follow the decreasing baseline density as a function of $\kperp$. 

The two simple shapes presented here are for frequency independent calibration errors. Frequency dependent calibration errors produce residuals that oscillate along the visibilities' lengths in $f$ and map almost directly into the $\kpar$ direction. It is hoped that many of these calibration errors will be slow and map to small $\kpar$, but they will generate additional foreground components that are not qualitatively described here. It is, however, straightforward to calculate their shapes. The kinds of errors that appear in the calibration $\B$ can be parametrized, and for each kind of error the residual power spectrum $\p(\k)$ can be explicitly calculated using Equation \ref{CalError}. This will be an important step for all 21~cm power spectrum measurements, but also is quite instrument dependent. Here we have concentrated on four power spectrum residuals that will be important for all upcoming observations.

%Intro discussing origin
%Figure showing effect

%\subsection{High spectral frequency component}

%part due to delta functions along f.
%identify in figure and Datta

%\subsection{Subtracting wrong PSF component}

%part due to PSF errors and/or angular band power (incoherent)
%identify in figure and Datta

%\subsection{Identification of calibration errors in k-space}

%Discuss identification of errors in k-space and k-f space. In particular the use of masks to identify errors in individual antenna calibrations and the coherent phase structure in k-space "image"
%

\section{Discussion}
\label{DisSec}

Using the results of \S 2--\S 4 we can quantitatively and qualitatively understand the origin of some of the different wedge-shaped structures seen in advanced simulations. This understanding also allows us to invert the process and associate observed power spectrum shapes with specific calibration and foreground subtraction errors. In practice, we feel it is this ability to identify the origin of observed power spectrum contaminations that will be the most influential result of this work. 

One of the key problems facing EoR observations is identifying calibration and foreground subtraction errors below the imaging limit. When you are working several orders of magnitude below the confusion limit, different subtraction or calibration algorithms can produce nearly identical images with very different power spectra. Further, when looking at measured power spectra in $\kpar$ vs.\ $\kperp$, to date it has been impossible to identify the causes of observed contamination. Using the developments in this article, not only can we compare the performance of these analyses, we can identify the specific sources of observed contamination. A few examples might include:
\begin{itemize}
  \item Two calibration algorithms that both produce good images can be compared, and their contribution to the power spectrum in the EoR window can be used to pick the superior algorithm (\S\ref{CalSec}).
  \item Symmetric bands of power as seen in Figure \ref{FigWedge} (accounting for projection), can be used to identify higher source subtraction errors near the field edge (\S\ref{AmpSec}).
  \item Linearly increasing bands of power can be used to identify pointing offsets in particular regions of the survey (e.g., far from a calibrator or near the ionospheric equator, \S\ref{LocationSec}).
  \item Systematic errors can be estimated by observing the contributions of specific foregrounds and their covariance with the EoR signal (\S\ref{FrameworkSec}, \citealt{Morales:2006p147}).
\end{itemize}

In this paper we have described four types of mode-mixing contamination any EoR instrument will see. The next paper in this series will explore the array-dependent effects and the influence of inhomogeneities in the baseline distribution (Hazelton et al.\ in preparation). Future papers will use these results to identify and quantify the mode-mixing contamination seen in 32 antenna MWA observations, and develop more advanced statistical methods to mitigate the contaminations due to calibration errors and array inhomogeneities. 
%In future work (Hazelton et al.\ in preparation, a) we use our new understanding to identify and quantify the sources of foreground contamination seen in observations with the 32 antenna MWA prototype.  and to suggest techniques for obtaining more robust EoR power spectrum estimates (Hazelton et al.\ in preparation, b)

\hspace{.5 cm}
\section*{Acknowledgements}

This work has been supported by the National Science Foundation Astronomy Division through CAREER award \#0847753 and NSF Postdoctoral Fellowship \#1003314, and by the University of Washington. We'd particularly like to thank Harish Vedantham, Udaya Shankar, Ravi Subrahmanyan, Judd Bowman for the discussions that initiated this work.

\bibliographystyle{apj}
\bibliography{./morales}

\begin{thebibliography}{}

\bibitem[\protect\citeauthoryear{Bowman, Morales, \& Hewitt}{Bowman
  et~al.}{2009}]{Bowman:2009p4044}
Bowman, J.~D., Morales, M.~F.,  \& Hewitt, J.~N. 2009, The Astrophysical
  Journal, 695, 183

\bibitem[\protect\citeauthoryear{Datta, Bowman, \& Carilli}{Datta
  et~al.}{2010}]{Datta:2010p4788}
Datta, A., Bowman, J.,  \& Carilli, C. 2010, The Astrophysical Journal, 724,
  526

\bibitem[\protect\citeauthoryear{Furlanetto, Oh, \& Briggs}{Furlanetto
  et~al.}{2006}]{Furlanetto:2006p341}
Furlanetto, S.~R., Oh, S.~P.,  \& Briggs, F.~H. 2006, Physics Reports, 433,
  181, Elsevier B.V.

\bibitem[\protect\citeauthoryear{Geil, Gaensler, \& Wyithe}{Geil
  et~al.}{2011}]{Geil:2011p4901}
Geil, P., Gaensler, B.,  \& Wyithe, J. 2011, Monthly Notices of the Royal
  Astronomical Society, 1, 1416

\bibitem[\protect\citeauthoryear{Gnedin \& Shaver}{Gnedin \&
  Shaver}{2004}]{Gnedin:2004p2873}
Gnedin, N.~Y.,  \& Shaver, P.~A. 2004, The Astrophysical Journal, 608, 611 (c)
  2004: The American Astronomical Society

\bibitem[\protect\citeauthoryear{Harker et~al.}{Harker
  et~al.}{2009}]{Harker:2009p4243}
Harker, G., et~al. 2009, Monthly Notices of the Royal Astronomical Society,
  397, 1138 (c) Journal compilation {\copyright} 2009 RAS

\bibitem[\protect\citeauthoryear{Hogg}{Hogg}{1999}]{Hogg:1999p923}
Hogg, D.~W. 1999, arXiv, astro-ph

\bibitem[\protect\citeauthoryear{Jeli{\'c} et~al.}{Jeli{\'c}
  et~al.}{2008}]{Jelic:2008p4524}
Jeli{\'c}, V., et~al. 2008, Monthly Notices of the Royal Astronomical Society,
  389, 1319

\bibitem[\protect\citeauthoryear{Liu \& Tegmark}{Liu \&
  Tegmark}{2011}]{Liu:2011p4789}
Liu, A.,  \& Tegmark, M. 2011, Physical Review D, 83, 103006

\bibitem[\protect\citeauthoryear{Liu, Tegmark, \& Zaldarriaga}{Liu
  et~al.}{2009}]{Liu:2009p4716}
Liu, A., Tegmark, M.,  \& Zaldarriaga, M. 2009, Monthly Notices of the Royal
  Astronomical Society, 394, 1575

\bibitem[\protect\citeauthoryear{Matteo, Ciardi, \& Miniati}{Matteo
  et~al.}{2004}]{DiMatteo:2004p2755}
Matteo, T.~D., Ciardi, B.,  \& Miniati, F. 2004, Monthly Notices of the Royal
  Astronomical Society, 355, 1053 (c) 2004 RAS

\bibitem[\protect\citeauthoryear{Matteo et~al.}{Matteo
  et~al.}{2002}]{DiMatteo:2002p2793}
Matteo, T.~D., Perna, R., Abel, T.,  \& Rees, M.~J. 2002, The Astrophysical
  Journal, 564, 576 (c) 2002: The American Astronomical Society

\bibitem[\protect\citeauthoryear{McQuinn et~al.}{McQuinn
  et~al.}{2007}]{McQuinn:2007p220}
McQuinn, M., Lidz, A., Zahn, O., Dutta, S., Hernquist, L.,  \& Zaldarriaga, M.
  2007, Monthly Notices of the Royal Astronomical Society, 377, 1043

\bibitem[\protect\citeauthoryear{McQuinn et~al.}{McQuinn
  et~al.}{2006}]{McQuinn:2006p222}
McQuinn, M., Zahn, O., Zaldarriaga, M., Hernquist, L.,  \& Furlanetto, S.~R.
  2006, The Astrophysical Journal, 653, 815 (c) 2006: The American Astronomical
  Society

\bibitem[\protect\citeauthoryear{Morales \& Wyithe}{Morales \&
  Wyithe}{2010}]{Morales:2010p4786}
Morales, M.,  \& Wyithe, J. 2010, Annual Review of Astronomy and Astrophysics,
  48, 127

\bibitem[\protect\citeauthoryear{Morales, Bowman, \& Hewitt}{Morales
  et~al.}{2006}]{Morales:2006p147}
Morales, M.~F., Bowman, J.~D.,  \& Hewitt, J.~N. 2006, The Astrophysical
  Journal, 648, 767 (c) 2006: The American Astronomical Society

\bibitem[\protect\citeauthoryear{Morales \& Hewitt}{Morales \&
  Hewitt}{2004}]{Morales:2004p803}
Morales, M.~F.,  \& Hewitt, J.~N. 2004, The Astrophysical Journal, 615, 7 (c)
  2004: The American Astronomical Society

\bibitem[\protect\citeauthoryear{Morales \& Matejek}{Morales \&
  Matejek}{2009}]{Morales:2009p4730}
Morales, M.~F.,  \& Matejek, M. 2009, Monthly Notices of the Royal Astronomical
  Society, 400, 1814

\bibitem[\protect\citeauthoryear{Oh \& Mack}{Oh \& Mack}{2003}]{Oh:2003p2809}
Oh, S.~P.,  \& Mack, K.~J. 2003, Monthly Notices of the Royal Astronomical
  Society, 346, 871

\bibitem[\protect\citeauthoryear{Santos, Cooray, \& Knox}{Santos
  et~al.}{2005}]{Santos:2005p2174}
Santos, M.~G., Cooray, A.,  \& Knox, L. 2005, The Astrophysical Journal, 625,
  575 (c) 2005: The American Astronomical Society

\bibitem[\protect\citeauthoryear{Tegmark}{Tegmark}{1997}]{Tegmark:1997p2012}
Tegmark, M. 1997, Physical Review D, 55, 5895 (c) 1997: The American Physical
  Society

\bibitem[\protect\citeauthoryear{Vedantham, Shankar, \& Subrahmanyan}{Vedantham
  et~al.}{2011}]{Vedantham:2011p4902}
Vedantham, H., Shankar, N.~U.,  \& Subrahmanyan, R. 2011, arXiv, astro-ph.IM

\bibitem[\protect\citeauthoryear{Wang et~al.}{Wang
  et~al.}{2006}]{Wang:2006p503}
Wang, X., Tegmark, M., Santos, M.~G.,  \& Knox, L. 2006, The Astrophysical
  Journal, 650, 529 (c) 2006: The American Astronomical Society

\bibitem[\protect\citeauthoryear{Zaldarriaga, Furlanetto, \&
  Hernquist}{Zaldarriaga et~al.}{2004}]{Zaldarriaga:2004p227}
Zaldarriaga, M., Furlanetto, S.~R.,  \& Hernquist, L. 2004, The Astrophysical
  Journal, 608, 622 (c) 2004: The American Astronomical Society

\end{thebibliography}

\end{document}